\documentstyle[multicol,aps,prl,epsfig]{revtex}

\begin{document}
\title{Revisiting ''swings'' in the crossover features of Ising thin films near $%
T_{c}(D)$}
\author{Manuel I. Marqu\'es and Julio A. Gonzalo}
\address{Departamento de F\'{i}sica de Materiales C-IV\\
Universidad At\'onoma de Madrid\\
Cantoblanco 28049 Madrid Spain\\
email: julio.gonzalo@uam.es}
\author{Juan Romero and Luis F. Fonseca}
\address{Department of Physics\\
University of Puerto Rico\\
San Juan, PR, 00931}
\date{February 15, 2000}
\maketitle

\begin{abstract}
''Swing'' effects at the onset of crossover towards two dimensional behavior
in thin Ising films are investigated close to $T_{c}(D)$ by means of Monte
Carlo calculations. We find that the effect is extremely large for the
specific heat effective critical exponent, in comparison with the ''swing''
already noted by Capehart and Fisher for the susceptibility. These effects
change considerably the system's evolution with thickness $(D)$ from
two-dimensional to three-dimensional behavior, forcing the effective
exponents to pass near characteristic Tri Critical Point (TCP) values.
\end{abstract}

\pacs{PACS numbers: 75.10.H, 64.60.F, 75.70, 02.50.N}



\begin{multicols}{2}
\narrowtext
\parskip=0cm

Basic features of phase transitions in systems with thin film geometry have
been connected with the problem of the crossover from classical to quantum
transitions. The change from classical to quantum character of the
transition can be mapped to the evolution with thickness of the phase
transition in films (se f.i.\cite{Suzuki,Kogut}).That is the reason why a
detailed study of phase transitions in films may be particularly useful for
the study of quantum phase transitions apart from the intrinsic usefulness
of studying changes in systems with a few layers of thickness. The effective
critical exponents and the evolution near the critical point has been
extensively studied by means of series expansions \cite{Capehart}, the
renormalization group \cite{Oconnor}, and Monte Carlo calculations in Ising
systems \cite{Schielbe}, as well as in the X-Y model \cite{Janke}. For
systems with thin film geometry, the correlation length is much smaller than
the film thickness $(D),$ sufficiently below and above the critical point
(i.e. relatively far from $T_{c}(D)$). Once the correlation length grows
sufficiently (i.e. close to $T_{c}(D)$) the system notices that its critical
behavior cannot be that of a three-dimensional system and the crossover to
the two-dimensional behavior begins. From the point of view of the {\bf %
effective critical exponents} this means that the system is initially
evolving towards {\bf three-dimensional} behavior until a crossover to {\bf %
two-dimensional} behavior takes place. The film thickness can be
characterized by the value of the effective critical exponents just at the
onset of this crossover.

The pioneering work of Capehart and Fisher \cite{Capehart} noted that for
the case of the effective critical exponent corresponding to the
susceptibility ($\gamma _{eff}$) an {\bf ''under-swing''} behavior was
apparent just before the crossover. This characteristic behavior means that
for a certain thickness $D^{\ast }$ the effective critical exponent reaches
a minimum with a value $\gamma _{m}(D^{\ast })<\gamma ^{3D}<\gamma ^{2D}$.\
This kind of behavior was attributed to surface effects, due to the lower
value of the thickness ($D<<L$). In principle one might expect to find an
enhancement of the pehenomenon using {\bf free} boundary conditions in
comparison with {\bf periodic} boundary conditions, as indeed it was seen
the case.

Since that time there has not been much work on the problem considered
because research has been devoted strictly to the very close vicinity of the
critical point. Monte Carlo simulations \cite{Schielbe} have shown the
existence of this ''under-swing'', but no attempt has been made to
characterize this phenomenon. In principle this ''under-swing''is a small
effect, since $\gamma _{m}(D^{\ast })$ is close to $\gamma ^{3D}$, but
several very interesting questions can be asked concerning this phenomenon:
a) We know that there should be a value of the thickness $D$ for which this
effect should be maximum, $D^{\ast },$ because eventually $\gamma _{eff}$
must increase again as $(D\rightarrow L)$ towards $\gamma ^{3D}$: What is
the value $D^{\ast }$ of this characteristic thickness? b) Is it possible to
get more pronounced ''swing'' effects in other critical exponents?, c) What
are the values of these effective critical exponents for $D^{\ast }$
corresponding to the maximum ''swing''? d) Is there a substantial difference
between the exponent values obtained using {\bf periodic} and {\bf free}
boundary conditions?

In the present work we will address these questions studying the thickness
dependence of the {\bf effective critical exponents} ($\beta _{eff},$ $%
\gamma _{eff},$ $\delta _{eff},$ $\alpha _{eff}$) of \ Ising film ($L\times
L\times D$), describing the evolution from the pure two-dimensional Ising
system ($D=1$) towards the three-dimensional system ($D=L$) system. In order
to obtain the actual behavior of the effective critical exponents we will
make use of the fact that the scaling relations hold all the way before and
all through the crossover region \cite{Oconnor,Marques}.

In the present work we report results on phase transitions in Ising plates
of equal area ($L=100$) and different thickness ($D=2,3,4,5,7,9,12)$ by
Monte Carlo calculations. In order to reduce the critical slowing down
effect near the critical point we use the Wolff single cluster algorithm 
\cite{Wolff}, with more than 50.000 MCS . To ensure equilibrium we start our
calculations with an early thermalization ($T\simeq 0K$ and $H=0$) and we
increase the temperature in very small (non-constant) steps as we get closer
and closer to the critical point. These very small temperature steps give
rise to large fluctuations in the numerical derivatives. We did smooth the
data taking derivatives including up to the 5$^{th}$ nearest neighboring
points. We obtain the evolution of the effective critical exponents $\beta
_{eff}$ and $\gamma _{eff}$ by a direct determination of the magnetization $%
M(T)$ and of the susceptibility $\chi (T)=\left\langle M^{2}\right\rangle
-\left\langle M\right\rangle ^{2}$ using the standard relations: 
\begin{equation}
\beta _{eff}=\frac{\partial logM(T)}{\partial log[T_{c}(D)-T]},\qquad \gamma
_{eff}=\frac{\partial log\chi (T)}{\partial log[T_{c}(D)-T]}
\end{equation}
The critical temperature $T_{c}(D)$ corresponding to each particular
thickness $D$ is obtained in the usual way by means of the Binder cumulant
method \cite{Binder} (see f.i. \cite{Schielbe,Marques,BinderII}).

Every calculation has been performed using {\bf free} and {\bf periodic}
boundary conditions. This comparison is important, because real films, due
to substrate effects, are not pure free-surface systems but have a mixture
of free and constrained surfaces. A direct comparison has been performed
elsewhere \cite{MarquesII} for the D dependence of the critical temperature.
Here we will carry out this comparison explicitly for the behavior of the
effective critical exponents.

We present in Fig.1a $\beta _{eff}$ vs. $log[T_{c}(D)-T]$ for $D=3,5,9$.
Three zones are clearly visible: (i) initial evolution towards the {\bf %
three-dimensional} value, (ii) {\bf crossover} zone towards the {\bf %
two-dimensional} value, and (iii) {\bf finite size} effects zone. Note that
for $D=3$ there is nearly no crossover, since the system is still almost
two-dimensional, and as $D$ increases, the maximum effective critical
exponent, defined just before the crossover starts, $\beta _{m}(D)$, grows
tending towards the three-dimensional value ($\beta ^{3D}\simeq 0.33$). We
remark the clear difference between exponents with {\bf periodic} and {\bf %
free} boundary conditions. Note how the exponent for free boundary
conditions always rises (for the same thickness) to a maximum which is
closer to the corresponding three-dimensional value than the same effective
critical exponent for periodic boundary conditions. Finally both (free and
periodic exponents) collapse together in the crossover. This exponent, as
will be seen also for $1/\delta _{eff}$ below, does not present any
''swing'' effect. It means that we do not find any value of $D$ for which $%
\beta ^{2D}<\beta ^{3D}<\beta _{m}(D).$
\begin{figure}
\epsfxsize=\columnwidth\epsfysize=15cm\epsfbox{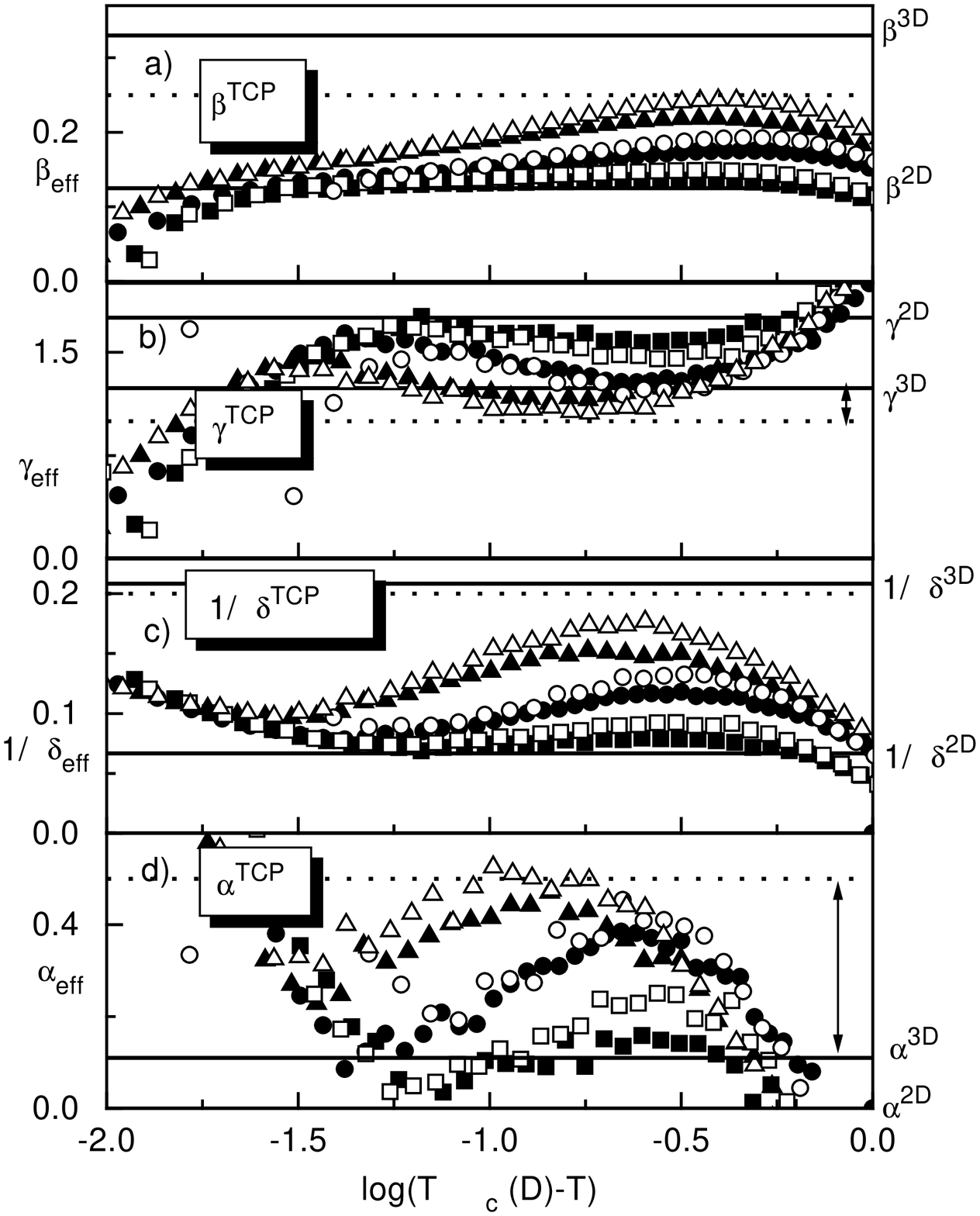}
\caption{Evolution of the effective critical exponents with temperature
for different thickness D=3 (squares), D=5 (circles) and D=9 (triangles)
with periodic (full) and free (open) boundary conditions. Two dimensional
and three dimensional critical exponents are marked (full lines) together
with the Tri Critical Point values (dashed line). The arrows indicate the
''under-swing'' (b) and ''over-swing'' (d) behavior.}
\label{fig1}
\end{figure}
The ''under-swing'' effect noted by Capehart and Fisher \cite{Capehart} is
explicit for the case of the susceptibility. In order to check this effect,
we present in Fig 1b results for $\gamma _{eff}$ vs. $log[T_{c}(D)-T]$ for $%
D=3,5,9.$ Note how the ''under-swing'' is clearly detectable for values of $%
D $ close to $D=9.$ This ''under-swing'' is visible not just for the free
boundary conditions, but also for the periodic boundary conditions as was
pointed out in Ref. \cite{Capehart}. This is the first time in our knowledge
that the ''under-swing'' effect is explicitly shown to exist under periodic
boundary conditions, where surface effects are reduced.

In order to get a more complete picture of the dependence with thickness of
the effective critical exponents we study also the effective critical
exponents $1/\delta _{m}(D)$ and $\alpha _{m}(D)$. As usual, $1/\delta
_{m}(D)$ should be obtained as the value just before $1/\delta _{eff}$
starts the crossover to the two-dimensional value, $1/\delta _{eff}$ may be
obtained making use of the scaling relation \cite{Yeomans} 
\begin{equation}
1/\delta _{eff}=\left(\frac{\gamma _{eff}}{\beta _{eff}}+1\right) ^{-1}
\end{equation}
As mentioned above, this relationship has been proven to hold, not just near
the critical point, but also at the crossover region, and before \cite
{Oconnor,Marques}. The results for $1/\delta _{eff}$ vs. $log[T_{c}(D)-T]$
are presented in Fig.1c. The behavior is very similar to the one observed
for $\beta _{eff}$. There is no ''over-swing''. Thus we do not have any
value of $D$ for which $1/\delta ^{2D}<1/\delta ^{3D}<1/\delta _{m}(D).$

The other interesting exponent to study is the effective specific heat
critical exponent. We are able to determine explicitly the evolution of this
effective critical exponent by means of the relation \cite{Yeomans}: 
\begin{equation}
\alpha _{eff}=2-2\beta _{eff}-\gamma _{eff}
\end{equation}
Fig 1d presents the results for $\alpha _{eff}$ vs. $log[T_{c}(D)-T].$ We
find and {\bf extremely enhanced ''over-swing''. }We find clearly that $%
\alpha ^{2D}<\alpha ^{3D}<\alpha _{m}(D)$, not just for $D=9$ but also for $%
D=5$ and $D=3$. Another interesting result is that the effect is very
clearly visible both for {\bf free} and for {\bf periodic} boundary
conditions.

We may note that in Fig1c and 1d the final data for $T\rightarrow T_{c}(D)$
are not representative, because they correspond to the finite size effects
region of $\beta _{eff}$ and $\gamma _{eff}$.

The best way to show the ''swing'' effect is perhaps to plot the values
obtained for $\alpha _{m}(D)$ vs. $D$ and for $\gamma _{m}(D)$ vs. $D$,
together with the results obtained for $\beta _{m}(D)$ and $\delta _{m}(D).$
These results are presented in Fig.2.\ Several points are made clear: a)
''Swing'' effects exist only for the specific heat $(\alpha _{eff})$ and the
susceptibility $(\gamma _{eff})$ effective critical exponents b) The
''swing'' is enhanced clearly in the effective heat exponent, f.i. we get a
ratio $\alpha _{m}(D=9)/\alpha ^{3D}\simeq $ $5$ while for $\gamma _{m}(D)$
we just find $\gamma ^{3D}/\gamma _{m}(D=9)\simeq 1.24.$ This means that the
effect that is relatively small in the susceptibility can not be ignored in
the specific heat c) The maximum ''swing'' effect is found for values of $D$
close to $10$, that is, we should take $D^{\ast }\approx 10.$ d) ''Swings''
appears for both, free and periodic boundary conditions, and are more
pronounced for the former.
\begin{figure}
\epsfxsize=7.25cm\epsfysize=13cm\epsfbox{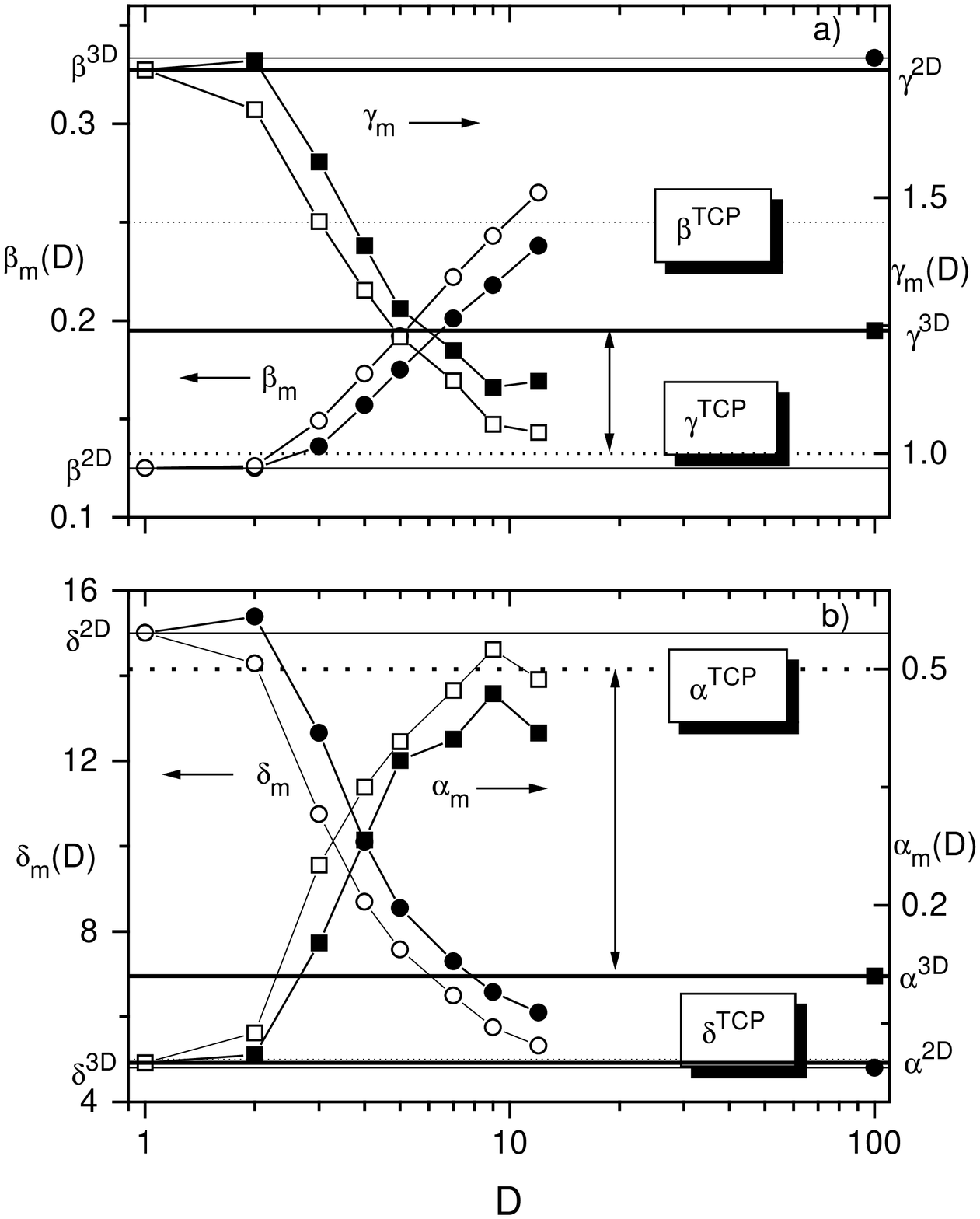}
\caption{Effective critical exponents at the onset of the crossover vs.
thickness for periodic (full) and free (open) boundary conditions. Two
dimensional and three dimensional critical exponents are marked (full line)
together with the Tri Critical Point value (dashed line). The arrows
indicate the ''under-swing'' in $\gamma _{eff}$ (a) and ''over-swing'' in $%
\alpha _{eff}$ (b).}
\label{fig2}
\end{figure}
Now we focus attention on the exponent values obtained for $D=9\approx
D^{\ast }$. The results for periodic and free boundary conditions are
presented in Table I. They are compared with the {\bf two-dimensional}
values, the {\bf three-dimensional} values and with the {\bf Tri Critical
Point} (TCP) values. As it is known, a Tri Critical Point is at the limit
separating {\bf continuous }(2$^{nd}$ order) from {\bf discontinuous} (1$%
^{st}$ order) transitions \cite{Gonzalo}. Note that the exponent values
corresponding to the Tri Critical Point are close to those for $D=9\approx
D^{\ast },$ with errors ranging from three to twelve percent. Clearly, the
evolution of the effective critical exponents [$\beta _{m}(D),\alpha
_{m}(D),1/\delta _{m}(D)$ and $\gamma _{m}(D)$] from the two-dimensional
values ($D=1$) to the three-dimensional values ($D=L)$ is not monotonous but
appears in all cases to come close to the respective Tri Critical Point
value for $D\approx D^{\ast }$. This effect is made more explicit in plots
of \ $\alpha _{m}(D)$ vs. $\gamma _{m}(D)$ and $1/\delta _{m}(D)$ vs. $\beta
_{m}(D)$ (see Fig.3a and 3b).
\begin{figure}
\epsfxsize=\columnwidth\epsfysize=10cm\epsfbox{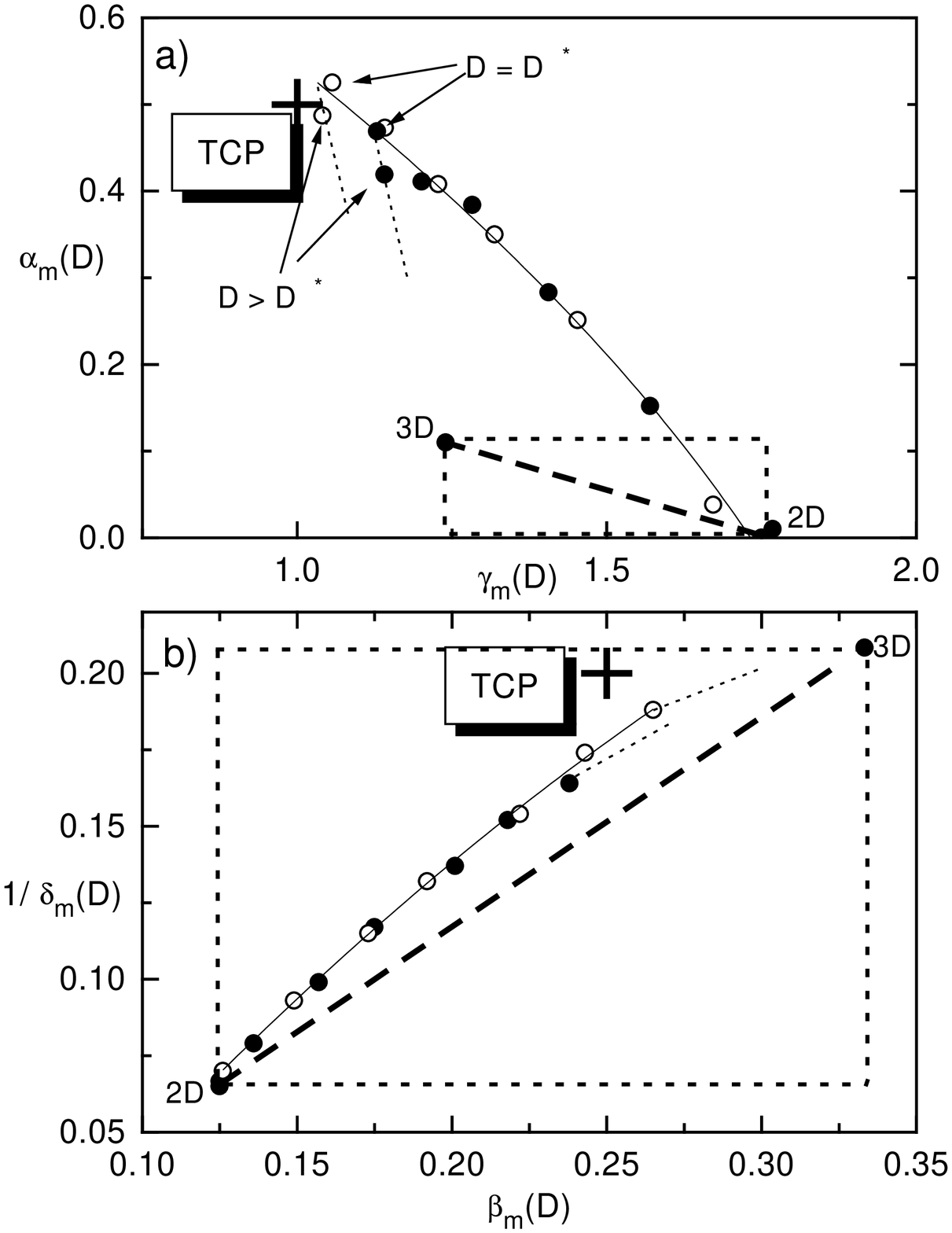}
\caption{Evolution of the effective critical exponents at the onset of
the crossover as the thickness of the system increases for periodic (full)
and free (open) boundary conditions. The straight dashed line indicates a
linear evolution, and the full line is a guide for the eye indicating the
observed evolution. Note that the effective exponents tend towards the Tri
Critical Point values (marked with an cross). Dotted lines indicate the
expected behavior towards the three-dimensional value.}
\label{fig3}
\end{figure}
Fig.3a shows that, in the case of $\alpha _{m}(D)$ vs. $\gamma _{m}(D),$the
evolution of the plot from $D<<D^{\ast }$ $[\gamma _{m}(D)\sim \gamma
^{2D},\alpha _{m}(D)\sim \alpha ^{2D}]$ onwards shows an spectacular turn
towards the {\bf Tri Critical Point} pair of values $(\gamma ^{TCP},\alpha
^{TCP}).$ Then for $D>D^{\ast }$ the evolution towards the pure
three-dimensional values begins. Note that the data points get away from the
box defined by $(\gamma ^{2D},\alpha ^{2D})\Longleftrightarrow (\gamma
^{3D},\alpha ^{3D}),$ making explicit the existence of ''swing effects''. An
interesting feature of our results is that the general behavior appears to
follow the same well defined line independently of the boundary conditions
used.

For the case of $1/\delta _{m}(D)$ vs. $\beta _{m}(D)$ the values
corresponding to a Tri Critical Point are also closely approached. The
non-existence of swing effects in this case is also explicit since the pair
of values $(\beta _{m}(D),1/\delta _{m}(D))$ do not leave the box.

In conclusion we have presented Monte Carlo data for the evolution of {\bf %
effective critical exponents} (note that these are ''transient'' exponents,
not assyntotic, critical exponents) in thin Ising films. In summary we have
shown that: a) ''Swing effects'' are specially enhanced for the specific
heat effective exponent, $\alpha _{m}(D)$, b) They appear very clearly for
both free and periodic boundary conditions and c) ''Swing effects'' force
the effective exponents to pass near exponent values corresponding to a Tri
Critical Point (for $D^{\ast }\simeq 10$) well before the evolution towards
the three-dimensional values begins.

Our work shows that ''swing'' effects must become patent especially in the
case of the specific heat for any boundary conditions. It would be very
interesting to check this point experimentally. This result rises also the
basic question of {\bf why} Tri Critical Point exponents $(\beta
=1/4,1/\delta =1/5,\gamma =1,\alpha =1/2)$ describe so well the behavior of
thin films at the onset of the crossover, for characteristic thicknesses of $%
D^{\ast }\simeq 10$.
\begin{table}
\caption{Effective critical exponents for Ising films at onset of crossover and $D\simeq D^{\ast }$}
\label{puretable}
\begin{tabular}{ccccc}
& $\beta $ & $\gamma $ & 1/$\delta $ & $\alpha $ \\ 
\hline
Two Dimensional & 0.125 & 1.75 & 0.066 & 0.000 \\
D=9 (periodic) & 0.218 & 1.13 & 0.152 & 0.469 \\ 
D=9 (free) & 0.243 & 1.06 & 0.174 & 0.525 \\ 
Tri Critical Point & 0.250 & 1.00 & 0.200 & 0.500 \\ 
Three Dimensional & 0.330 & 1.24 & 0.208 & 0.110
\end{tabular}
\end{table}
We thank P.A.Serena for computing facilities and we acknowledge financial
support from DGCyT through grant PB96-0037.


\end{multicols}

\end{document}